\documentclass[a4paper]{article}
\usepackage[margin=1in]{geometry} 
\usepackage{amsmath}
\usepackage{amsthm}
\usepackage{amssymb}
\usepackage[utf8]{inputenc}
\usepackage{xcolor}
\usepackage{url}
\usepackage{hyperref}
\hypersetup{
 colorlinks,
 citecolor=blue,
 linkcolor=blue,
 urlcolor=blue
}
\usepackage{titlesec}
\usepackage{siunitx} 
\usepackage[comma,compress]{natbib}
\usepackage{graphicx, color}
\usepackage{algorithm, algpseudocode} 
\usepackage{mathrsfs}
\usepackage[font=small,skip=5pt]{caption}

\title{On the Need for a Near-Earth Object Characterization Constellation in Low-Earth Orbit}
\author{Nathan Golovich\thanks{golovich1@llnl.gov} \\ \\ Lawrence Livermore National Laboratory, 7000 East Avenue, Livermore, CA 94550, USA.}

\date{\today}
\begin{document}
\maketitle
\begin{abstract}
In 2005, the United States Congress passed a bill mandating the detection, tracking, cataloguing and characterization of 90\% of the 140 meter and larger near-Earth objects (NEOs) by 2020. At the deadline $\sim$35\% were detected, tracked and catalogued, but only a small fraction were characterized. At the present rate, it will take 40 years to meet the detection mandate, and there are insufficient global facilities dedicated to NEO characterization to come close to the characterization threshold. The major surveys focus mainly on detection and initial orbit determination, which must be refined in order to fully be tracked and catalogued. Characterization requires observations spanning multiple wavelengths, cadences, and instruments, so it is challenging for observers to acquire the requisite data in a timely manner for planetary defense. Two upcoming surveys will easily meet the 90\% threshold for detection, but each will require separate facilities to tip and queue to refine orbits and characterize new discoveries, and they will provide too many discoveries for ground and space-based assets to keep up with. Here, I argue for a constellation of proliferating small satellites carrying visible and infrared sensors that would offer the needed coverage and flexibility to follow up detections from current and upcoming surveys in a timely manner. Such a constellation would enable NASA to move beyond the detection focused investments and fully meet the 2005 Congressional mandate. 
\end{abstract}

\section{Introduction}

In the past few years, satellites in low-Earth orbit (LEO)\footnote{LEO generally implies roughly circular orbits ($e<0.25$) with altitudes below 2000 km.} have proliferated. These serve many purposes including scientific, civil, and commercial. The explosion in numbers has occurred due to the confluence of two primary factors:

\begin{itemize}
    \item Small satellite platforms have gotten cheaper due to competition and economies of scale. Off-the-shelf and space-ready options exist for satellite buses, communications systems, on-board processors, science instrumentation, cooling systems, attitude control, and other components. 
    \item The introduction of commercial space-launch companies has opened the door to ride-sharing, which has vastly decreased the cost to get to space \citep{Jones2018TheRL}.\footnote{The cost to LEO was $20\times$ less for a SpaceX Falcon 9 compared with NASA's space shuttle in 2011.}
\end{itemize}

\begin{figure}[htb!]
  \centering
  \includegraphics[width=0.66\textwidth]{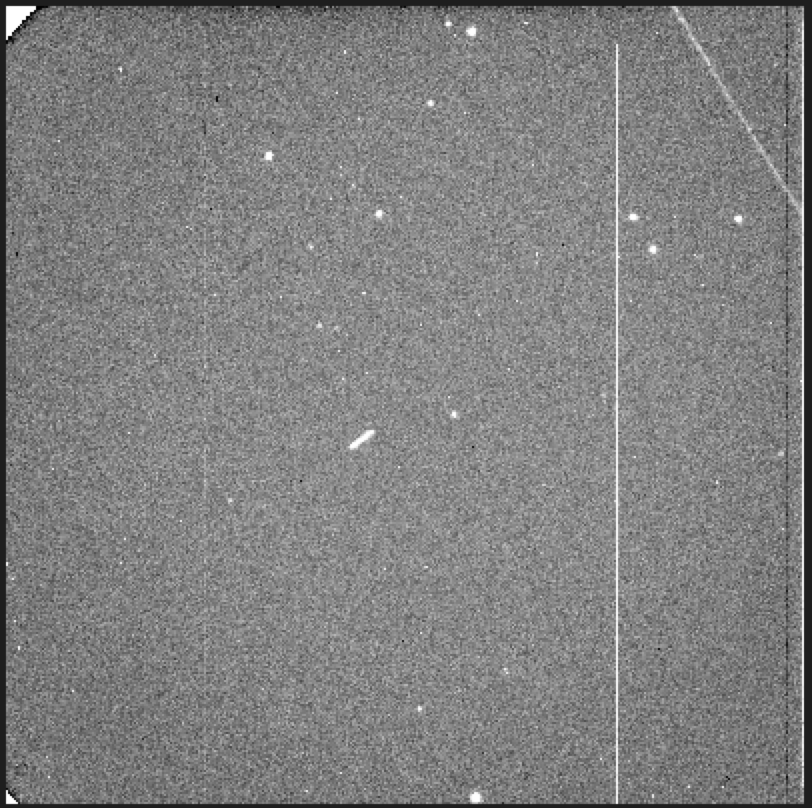}
  \caption{Recent Nickel telescope image showing satellite streaking through the top-right quadrant of the camera during the exposure time. Meanwhile, a NEO streaks through the center of the camera with a much smaller angular speed. Note that the vertical lines are dead pixel columns, not streaking objects.}
  \label{fig:1}
\end{figure}

Indeed, the number of satellites orbiting in LEO worries ground-based astronomers. Satellites are an ever-present contaminant in ground-based images (Figure \ref{fig:1} shows a recent example from Lick Observatory). This concern is due to not only the assumption of continued proliferation of LEO satellites, but also the increase in LEO satellite size (and thus brightness). Starlink and similar constellations are bright enough to saturate typical photometric images taken by professional telescopes. Numerous studies in the recent literature have explored the impact to ground based astronomy caused by large LEO constellations \citep[e.g.;][]{2020AA...636A.121H,2020AJ....160..226T, 2022AJ....163...21L, 2022ApJ...924L..30M}. The general consensus is that wide field surveys from large apertures could have as many as a half of their images compromised in the first and last hour of observing each night. Such surveys, unfortunately, are typical of surveys for near-Earth objects (NEOs\footnote{NEOs can be either asteroids or comets, but are predominantly asteroids. I will use NEO and asteroid interchangeably.}), which invariably have large etendue (product of aperture and field of view). 

Nevertheless, large ground-based observatories dedicated to detecting NEOs discover the vast majority of NEOs today. This will be true through the 2020s. In the 2030s, major ground and space based surveys will discover more faint NEOs than ever before. The two major next generation surveys will stress our ability to follow up NEOs for vital tracking and characterization studies. In this paper, I will assess the need for and benefits of a network of optical\footnote{I will use optical and visible interchangeably to describe light in the $\sim400-800$ nm range.} and near-infrared (NIR) sensors operating in space for planetary defense. The impending deluge of new NEO discoveries by next-generation survey instruments will overwhelm current assets, and such a constellation in LEO would be able to tip and queue the satellites to cover a larger range of orbital phase space, which would deliver the oft-overlooked characterization mission for NEOs. Tracking and characterizing NEOs quickly after discovery offers timely and vital information for planetary defense mitigation strategies, especially as more NEOs are discovered in the future that could have a higher probability of impacting Earth on a shorter timeline than has been typical. 

Such an occurrence happened following the 2004 discovery of 99942 Apophis, which was assessed to have a few percent chance of impacting Earth in 2029 or 2036 \citep{2005CBET..212....1G, 2006IAUS..229..215C}. Ultimately, later observations refined the orbit and retired the risk \citep[see e.g.,][]{2008Icar..193....1G, 2009DPS....41.4306C}. However, the experience led numerous astronomers to study the asteroid carefully for features that could inform planetary defense mitigation strategies \citep{2006SASS...25..168B, 2007IAUS..236..451C, 2007Icar..188..266D, 2007Icar..192..460R, 2009Icar..200..480B, 2011Icar..211..511Z, 2014Icar..233...48P, 2014AA...566A..22M, 2015MNRAS.449.2489B, 2018Icar..300..115B, 2018AJ....155..140R}. Simultaneously, numerous studies developed mitigation strategies rapidly after discovery \citep{2006AcAau..58..662M,2006Icar..182..482K, 2006AcAau..59..294I,2007IAUS..236..477B, 2008cosp...37..810E, 2008ChJAS...8..399F, 2009JSpRo..46..134M, 2010AcAau..66.1506Q, 2010MsT..........2M, 2010CosRe..48..430H, 2013AcAau..90...72W, 2013AcAau..90..112F, 2014AcAau.105...40Z,  2021JASS...38..105K}. This experience is possible to occur again because new surveys will be discovering NEOs more rapidly than ever before. This will increase the probability of another scenario that requires quick orbital track refinement and characterization. Ultimately, the 99942 Apophis scenario was used for a planetary defense exercise conducted during the 2020--2021 close approach to Earth \citep{2022PSJ.....3..123R}.

Led by fact that upcoming surveys will increase the likelihood of another scenario like 99942 Apophis while ground-based tracking and characterization assets are insufficient to serve this task, I decided to discuss the challenge at hand in this paper. In \S\ref{sec:mandate}, I will discuss the language of the NASA Authorization Act of 2005, which mandated NASA to detect and characterize large NEOs in the first place. I will map the text of the bill onto a set of what I call NEO ``missions'' for planetary defense that must be satisfied in order to fully meet the mandate. In \S\ref{sec:detection}, I will highlight the major NEO detection surveys of the present and future. I will also discuss how these meet the various NEO missions and what gaps remain for follow-up instruments. In \S\ref{sec:followup}, I will discuss the assets available to astronomers for follow up observations of NEOs to fill the gaps in information that the surveys are ill equip to fill themselves. In \S\ref{sec:constellation}, I will discuss how a constellation of small LEO satellites offer a technically effective and feasible solution to fully meet the mandate set by Congress. Finally, in \S\ref{sec:conclusions}, I will discuss the key takeaways of this exercise and offer my conclusions. 

\section{Dissecting the Congressional Mandate} \label{sec:mandate}
Section 321 of the NASA Authorization Act of 2005 (Public Law No. 109-155\footnote{\url{https://www.congress.gov/109/plaws/publ155/PLAW-109publ155.pdf}}) contains the text originally written as the George E. Brown Near Earth Object Survey Act\footnote{\url{https://www.congress.gov/bill/109th-congress/house-bill/1022}} (hereafter NEO Survey Act), which mandates NASA to ``detect, track, catalogue, and characterize the physical characteristics of near-Earth objects equal to or greater than 140 meters in diameter in order to assess the threat of such near-Earth objects to the Earth.'' 

The NEO Survey Act is often discussed as a mandate related to detecting NEOs. However, this overlooks the additional aspects of the NEO Survey Act's text, which explicitly lay out a series of actions (namely the verbs detect, track, catalogue, and characterize). Taken collectively, these actions enable the full assessment of the threat to Earth posed by NEOs larger than 140 meters. I will refer to these verbs in terms of a corresponding ``mission'' with each working toward the full mandate laid out in the bill. NASA has funded various efforts over the past few decades that respond to these missions, but the clear priority has been in funding the detection mission.

\subsection{Size Threshold in the NEO Survey Act}\label{subsec:size_thresh}
The NEO Survey Act includes a size threshold of 140 meters and larger, which is a relatively arbitrary cut off. It corresponds to the size of a typical NEO with an absolute magnitude $H=22$ \citep[assuming an albedo of 0.15; see Bowell et al. p. 524-556 in][]{binzel1989asteroids, 1997Icar..126..450H}. The population of NEOs with $H<22$ has been estimated to contain around 25,000 objects \citep{2021PSJ.....2...12H}. There are many asteroids with albedo lower than 0.15, and since the NEO Survey Act was written in terms of size, it will be necessary to discover many additional NEOs with $H>22$ to fully meet the mandate. The absolute magnitude for an asteroid is equivalent to $V$ magnitude under the assumption that the asteroid is 1 AU from both the Sun and the Earth with zero phase angle\footnote{Note this is technically not physical.}.

An impact by a 140 meter asteroid would certainly be problematic. These are typically referred to as ``city killers,'' but objects as small as 20 meters can cause substantial damage on the ground. This was made apparent in 2013 when a $\sim$20 meter asteroid exploded in the atmosphere above Chelyabinsk, Russia. The 500 kiloton air-burst generated a shock wave that injured over 1000 people and damaged thousands of buildings \citep{2013Natur.503..238B}. This event led to revised estimates that shifted the residual risk substantially toward NEOs of 10--50 meters in diameter \citep{2013Sci...342.1069P}. The NEO size distribution generally follows a broken power-law form \citep[see e.g.,][]{2021PSJ.....2...12H}. While there are $\sim$30,000 NEOs 140 meters or larger, there at least a few million NEOs as large as the Chelyabinsk meteor or larger \citep{2018Icar..312..181G, 2021PSJ.....2...12H}. For the majority of this paper, I will focus on the NEOs mandated by Congress; however, where appropriate, I will discuss the implications of the large number of NEOs that are dangerous by not covered by the mandate.

\subsection{NEO Detection Mission}
Here I will briefly describe the NEO detection mission as it relates to the NEO Survey Act. Later in \S\ref{sec:detection}, I will discuss in more detail the major surveys that carry out this mission. Planetary defense starts with the detection (or discovery) mission. For a detailed review of the NEO detection mission, I refer readers to \citet{NAP25476}. One key finding of this report suggests that a combination of optical and IR systems should continue to be funded since neither modality offers full information necessary for planetary defense. Objects are generally detected by major surveys, which have cadences designed to recover the same object a few times in a short span to issue a discovery and initial orbit determination (IOD). In this respect, prioritizing funding for surveys makes sense. The detection list can be triaged based on the IOD; however, detections are typically made while NEOs are close and only recoverable for a short duration, so missing the opportunity to track and characterize any newly discovered NEO could mean years of waiting for another opportunity. 

\subsection{NEO Tracking and Cataloguing Missions} 
NEO discoveries are generally posted to the Minor Planet Center (MPC)\footnote{\url{https://minorplanetcenter.net/}}, which keeps a list in need of new discoveries in need of refinement for astronomers to observe until the orbit is sufficiently well-tracked. Meanwhile, the MPC and other cataloging services such as NASA’s Center for Near Earth Object Studies (CNEOS)\footnote{\url{https://cneos.jpl.nasa.gov}} and the European Space Agency’s Near Earth Objects Coordination Centre\footnote{\url{https://neo.ssa.esa.int}} maintain catalogs and offer ephemeris services and other observing tools that enable astronomers to plan follow-up observations that help satisfy both the tracking and characterization missions. Many amateur astronomers contribute meaningfully to the tracking mission by observing new discoveries after they are posted to the MPC.

\subsection{NEO Characterization Mission}
Of the roughly 10,000 NEOs 140 meters or larger discovered today, only a small fraction are fully characterized. This is because constraining the physical characteristics of NEOs is an expensive task. Parameters of interest include the size, shape, rotational, and material properties. In this subsection, I will primarily discuss the methods astronomers use to constrain these features, which are important for planetary defense mitigation strategies and impact threat assessments \citep[e.g.;][]{1998Natur.393..437A, 2016Icar..269...50B, 2017JGCD...40.2417F, 2019EPSC...13...20D, 2020AcAau.166..290D, 2020DPS....5251202B, 2021plde.confE.195G, 2021AcAau.188..367K}. Readers are also encouraged to review the United State's NEO Preparedness and Action Plan\footnote{ \url{https://www.nasa.gov/sites/default/files/atoms/files/ostp-neo-strategy-action-plan-jun18.pdf}}. 

In 2022, NASA carried out the first planetary defense experiment to deflect an asteroid's trajectory with a collision with the Double Asteroid Redirection Test \citep[DART;][]{2021PSJ.....2..173R}. The success or failure of such a mission in a real-world deflection attempt could very well require precise of knowledge of these characteristics listed above so that the maximal impulse may be imparted upon impact \citep{1998Natur.393..437A,2016Icar..269...50B}. Other possible mitigation strategies exist ranging from painting asteroids white \citep{2022JAnSc..69..941K} to successive impulsive impacts or even deflecting or destroying a NEO with a nuclear explosion \citep{2020AcAau.166..290D}. The methods to optimally defend against a likely impact depend strongly on the lead time as well as knowledge of the asteroids physical properties. Thus, it is important to consider the gaps in our present investments in meeting the mandate's full list of requirements. For the remainder of this subsection, I will discuss methods to constrain various NEO features following discovery. I will point out when one method is more constraining than another and also when one method is more expensive (usually in terms of telescope time but also in terms of instrument complexity or difficulty of obtaining the measurement). 

\subsubsection{Size}
The size typically refers to the diameter, but it can also refer to the mass, volume, or cross-sectional area. Spherical geometry is often assumed, but the vast majority of minor planets have far too little mass to self-gravitate into a sphere. Rocky objects must be $\sim$500 km in diameter to be spherical. The second largest asteroid in the Solar System (2 Pallas) has a mean diameter of $510$ km. It is slightly tri-axial with $c/a\approx0.79$ and $b/a\approx0.95$. Larger objects in the Solar System are all more spherical. 

Asteroid size is important for several aspects of planetary defense. First, the entire cross-sectional area of an asteroid is able to absorb radiative impulses \citep{2020AcAau.166..290D}. Second, the drag of an asteroid entering the atmosphere depends on this area. Similarly, the non-Keplarian components of asteroid motion through the inner Solar System are key contributors to both the sourcing and the long-term instability of NEOs. The YORP \citep[e.g.,][]{2000Icar..148....2R, 2007Sci...316..274T, 2007Sci...316..272L} and Yarkovsky \citep[e.g.,][]{2003Sci...302.1739C} effects cause asteroids to rotate and increase their semi-major axis due to effects with Solar radiation. Each of these is dependent on the size of the asteroid. 

Ideally, astronomers estimate size with resolved imagery. This is typically done with radar for asteroids; however, adaptive optics has enabled resolved imagery of NEOs with large aperture telescopes \citep{2022Icar..37414790R}. Alternatively asteroids can be resolved with interferometry observations \citep[see e.g.,][]{2011Icar..211.1007L}. For each of these methods, the number of asteroids available are few, and each method is expensive to carry out. Most practically, diameter can be constrained with thermal emission in the mid-IR (MIR) wavelengths (4--25 $\mu$m) to $\sim10\%$ accuracy \citep{2011ApJ...736..100M}; however, sensors that cover this wavelength range are expensive HgCdTe technology, so these cameras have not been proliferated for rapid follow up purposes. They have been used for surveys of NEOs from space, which I will discuss in detail in \S\ref{sec:detection}.

\subsubsection{Mass}
Mass is among the most important measure of an asteroid's size for planetary defense since most mitigation strategies involve deflecting the asteroid's trajectory with an impulse, and the ability to deflect the asteroid directly depends on the mass; however, it is challenging to estimate the mass based on size estimates since the density of asteroids span a wide range. It is trivial to estimate the mass of a parent asteroid in a binary with a large mass ratio, but these are uncommon for smaller NEOs, and the secondary for a smaller NEO becomes more difficult to observe in such a scenario. Mass can be estimated for a large asteroid by the impact or close encounter of a smaller asteroid \citep{2022EPSC...16.1081S} or a man-made satellite \citep{2022PSJ.....3..237K}. Simulation work suggests that the mass can be estimated for loosely bound ``rubble-pile'' NEOs that fly by Earth \citep{1998Icar..134...47R}. However, efforts to measure the mass of asteroids are not common since the mass is less interesting scientifically than the density, which better probes material properties and constrains important questions of the formation of the Solar System.

\subsubsection{Material Properties}\label{sssec:material}
\begin{figure}[htb!]
  \centering
  \includegraphics[width=0.66\textwidth]{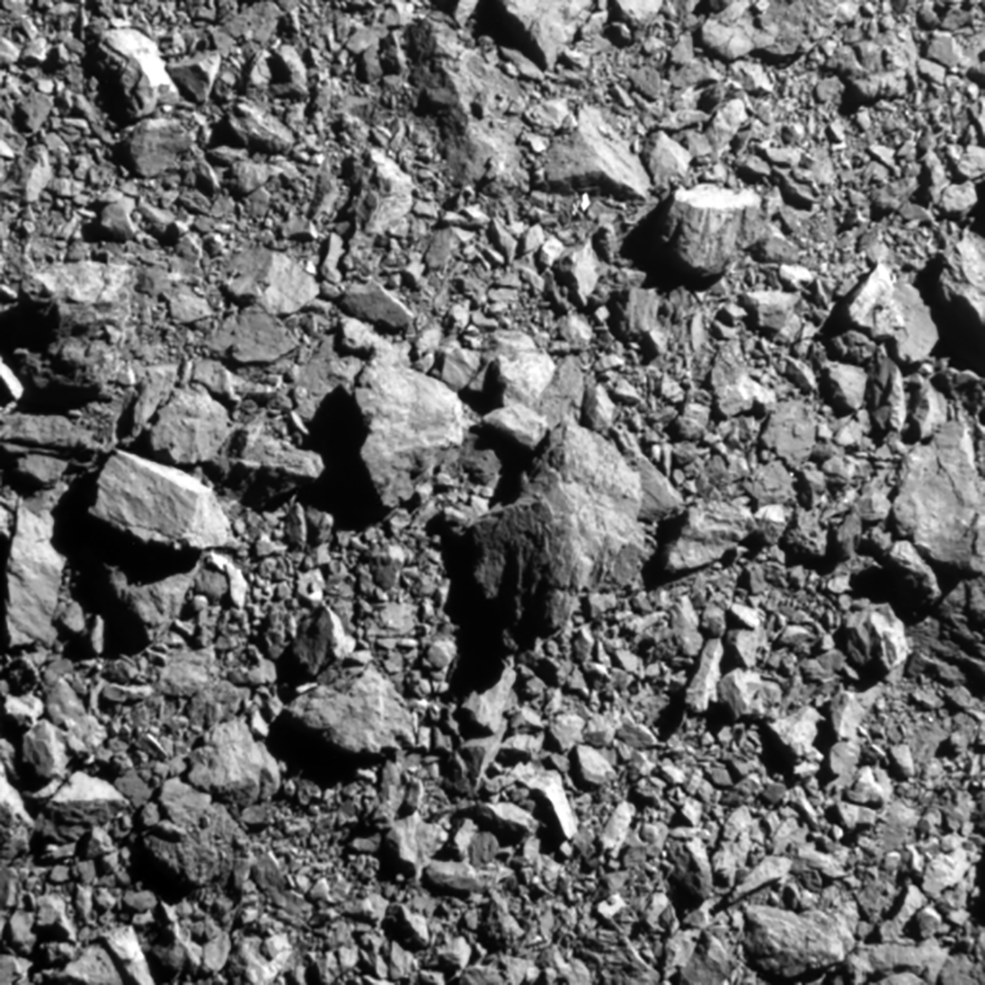}
  \caption{Final image taken by the DART mission before impact showing the material properties of the surface of Dimorphos from 7 km. The image shows a patch of the surface approximately 100 meters on a side. The largest boulders are $\sim$15 meters across, but there is a mix of sizes and regolith filling the gaps between large boulders. Image credit: NASA/Johns Hopkins APL. This and more images are available at \url{https://www.nasa.gov/feature/dart-s-final-images-prior-to-impact}.}
  \label{fig:2}
\end{figure}

The material properties of an asteroid are the most difficult to constrain with remote sensing. Material properties generally refer to chemical composition, but also composite nature, density, bulk strength, and albedo among other features. The most direct method to constrain the chemical composition with remote sensing is spectroscopy, which reveals absorption lines of chemicals on the asteroid; however, these are only indicative of the surface layer, which is often covered in a layer of fine regolith and mixed sizes from dust to boulders (see Figure \ref{fig:2}).

The interior of asteroids are much more difficult to constrain. There have been a three sample return missions of NEOs, but each only extracted material from the surface layer. The Japanese Aerospace Exploration Agency rendezvoused with 25143 Itokawa and 162173 Ryugu with their Hayabusa and Hayabusa2 missions, respectively. Most recently, NASA's OSIRIS-REx mission explored 101955 Bennu and extracted material, which will be returned later in 2023. Hayabusa's target 25143 Itokawa was shown to be an S-type asteroid similar to ordinary chrondite meteorite samples. It is a contact binary with two lobes with different densities, and it was shown to have water present \citep{2006Sci...312.1344A, 2006Sci...312.1330F, 2008AA...488..345D, 2011Sci...333.1113N, 2018LPI....49.1670J}. Its surface seemed to be exposed to space weathering for eight million years suggesting it was originally part of a larger asteroid and had broken off and recombined \citep{2006Sci...312.1344A}. Hayabusa2's target, 162173 Ryugu, was shown to have a similar surface age as 25143 Itokawa; however, it is more similar to C1 chrondite meteorites and is a Cb-type asteroid. The surface was shown to contain water, and the asteroid has a low density and uniform spectral features covering the surface. Its low density and porous surface boulders suggest it was formed from the recombination of loose material \citep{2019Sci...364..252S, 2019NatAs...3..971G, 2019Sci...365..817J, 2019Icar..331..179M}. 

The young age of the surface layers suggests that some NEOs formed recently compared with the age of the Solar System with the common picture being that they are formed from various escape routes from the asteroid belt \citep{1979Icar...37...96W, WISDOM198351, 2000Sci...288.2190B, 2002aste.book..409M, 2008Natur.454..858V, 2016Icar..268..340C}. NEOs are stable in the inner Solar System for a few million years \citep{2002Icar..156..399B, 2002aste.book..409M, 2016Natur.530..303G}. However, they also weather on timescales $<1$ Myr \citep{2009Natur.458..993V}. Space weathering refers to an array of processes that alter the surfaces including high energy particles from the Sun and cosmic rays as well as impacts from micrometeoroids. Some NEOs also appear to have unweathered surfaces, which has been explained by a number of processes including planetary encounters, YORP spin-up, small impacts, and thermal degradation. \citet{2023Icar..38915264D} showed that none of these processes can explain all trends observed in Q-type asteroids in the inner Solar System. A few broad trends were observed including fewer fresh surfaces on smaller asteroids, which could be explained by smaller asteroids being more likely to have higher bulk strength.

\begin{figure}[h!]
  \centering
  \includegraphics[width=\textwidth]{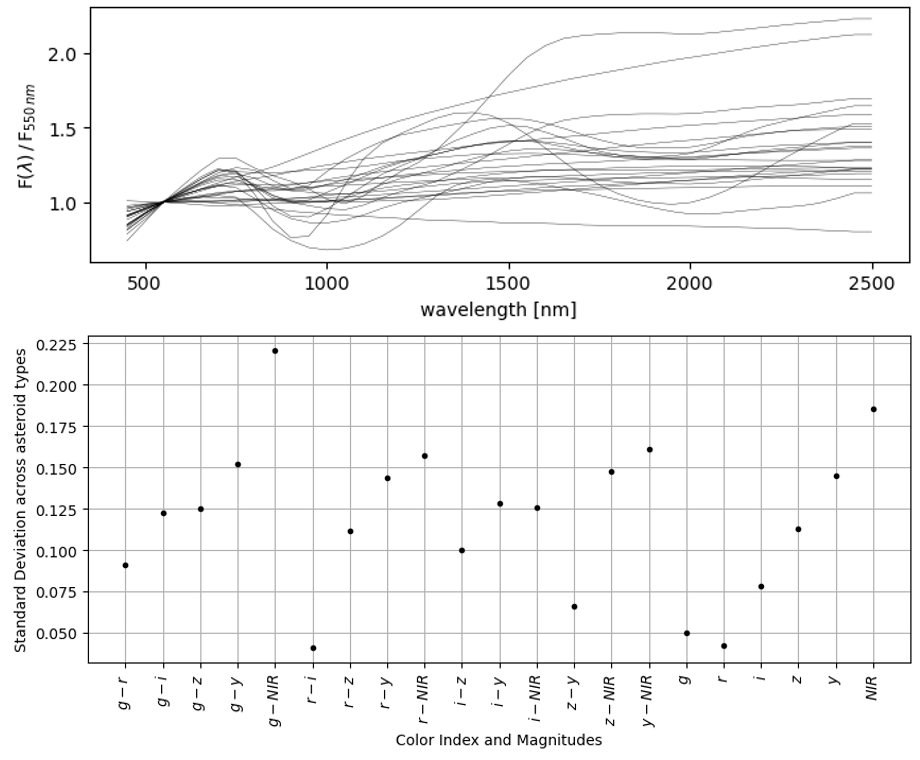}
  \caption{\emph{Top:} SEDs for all 24 asteroid types from \citet{2009Icar..202..160D}. The flux in each curve is normalized to the flux at 550 nm. There is a much more prominent spread in the NIR than visible wavelengths. The NIR colors are estimated assuming the quantum-efficiency curve for a InGaAs sensor unfiltered. \emph{Bottom:} The standard deviation of color measurements made with Rubin/LSST filters (omitting the u-band filter, which is bluer than the SEDs above) and an unfiltered NIR quantum efficiency curve. For each color index, the two relevant broad-band filters were convolved with all 24 spectral energy distributions and apparent magnitudes were estimated assuming a nominal flux and the standard deviation is reported.}
  \label{fig:3}
\end{figure}

However, since rendezvous and sample return missions are not economically feasible for more than a handful of NEOs, spectral studies of asteroids often reveal the most information we can obtain for the chemical structure of NEOs. \citet{2009Icar..202..160D} provides the canonical asteroid spectral classification system based upon visible to NIR spectroscopy of a few hundred asteroids. The 24 spectral classes are shown in the top panel of Figure \ref{fig:3} with the relative flux normalized at 550 nm. Many of these spectral classifications have been linked to various meteorite samples \citep{2022Icar..38014971D}. Furthermore, many of these spectral classes have also been linked to escape mechanisms from the asteroid belt \citep{2022AJ....163..165M}. These studies constitute an important line of work that will continue to increase the usefulness of spectral classification of NEOs for planetary defense purposes. 

However, spectroscopy across these wavelengths are difficult to measure for the majority of NEOs, especially rapidly after detection by large aperture survey instruments, which are likely to be detected further away. Since spectroscopy requires binning photons by wavelength, longer exposure times are needed to obtain sufficiently high signal-to-noise ratio (SNR) measurements to identify spectral features and classify asteroids. An alternative is multi-band photometry covering the visible to NIR wavelengths to coarsely resolve the spectral energy distribution (SED). A number of broad-band filters are in common use that span these wavelengths. Such a measurement is analogous to using broad-band filters to estimate the redshift of galaxies used for extra-galactic astrophysics and cosmology as opposed to the more expensive spectroscopic redshift measurements. In fact, most modern photometric filter sets are designed with this science in mind. A useful endeavor for planetary defense and Solar System science would be to design a filter set that maximizes the ability to approximate spectral classification with broad-band filters. 

In the bottom panel of Figure \ref{fig:3}, I show the variability in visible and NIR colors across the 24 spectral classifications. These were determined using the filters that will be used for the Vera C. Rubin Observatory Legacy Survey of Space and Time \citep[hereafter referenced together as Rubin/LSST or seperately referring to the telescope or the survey, respectively;][]{2019ApJ...873..111I} plus an unfiltered InGaAS NIR camera. Colors involving the NIR filter are most constraining. This method has been used in the literature \citep{2013AA...555L...2D} to identify a NEO as an L-type asteroid during a close approach, which is a good approximation of rapid characterization after discovery.

\subsubsection{Rotational State}
\begin{figure}[t!]
  \centering
  \includegraphics[width=\textwidth]{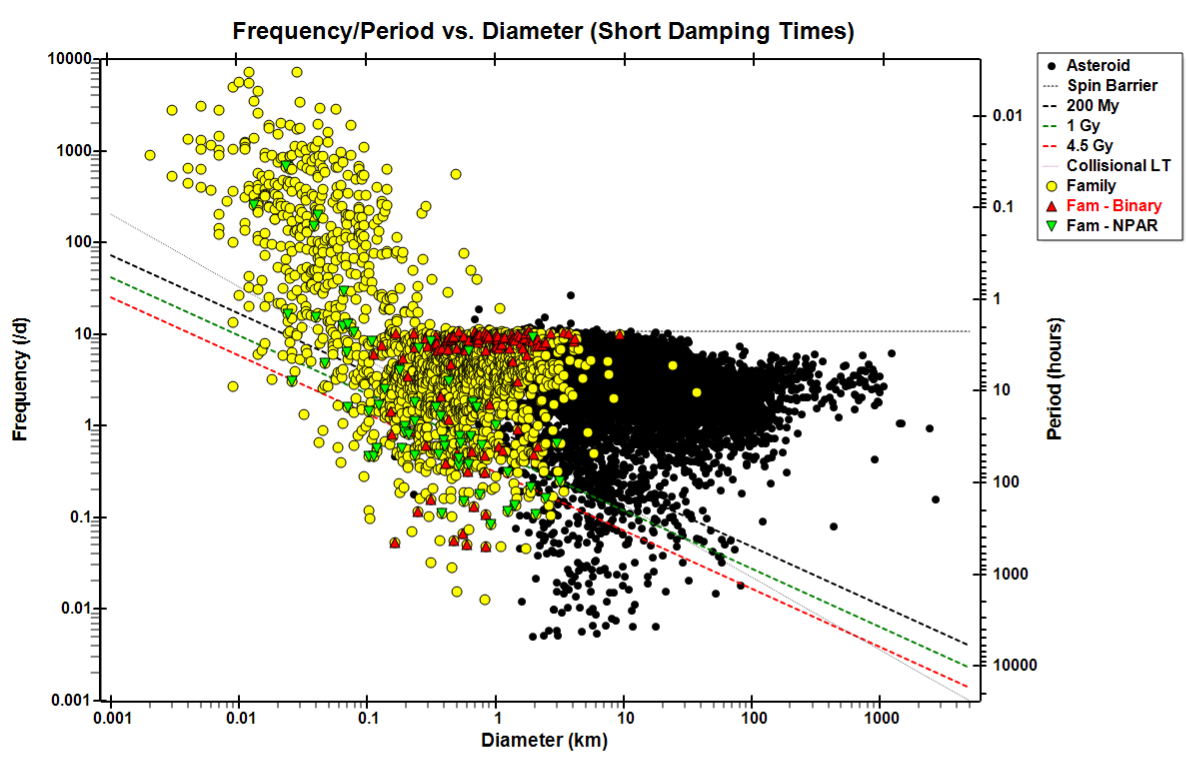}
  \caption{Rotation frequency versus diameter for known asteroids. The plot is made from data from the Asteroid Lightcurve Database available at \url{https://minplanobs.org/mpinfo/php/lcdb.php}. The black points are all asteroid in the database. Yellow points are NEOs with red and green points denoting binary NEOs and tumbling (NPAR refers to non-principle axis rotation), respectively. Smaller asteroids are more likely to be NEOs given their proximity.}
  \label{fig:4}
\end{figure}

Similar to asteroid size measurement, asteroid rotation is most easily constrained with a series of resolved images; however, similarly, this is expensive compared with a more practical and proliferating approach made available by optical photometry generated light curves. The period of rotation is observable in a light curve as the period of a repeating pattern. Asteroids have a wide range of rotational frequencies that form a peculiar pattern when looked at as a population as a function of diameter \citep[see Figure \ref{fig:4}, prepared by the Asteroid Lightcurve Database][]{2009Icar..202..134W}. This is referred to as a spin barrier, which is interpreted as the maximum frequency an asteroid can have before its gravitational strength is overcome by centrifugal forces caused by rotation. This suggests that solid asteroids larger than a few hundred meters in diameter are extremely rare. The common interpretation is that large asteroids are ``rubble piles'' loosely bound by gravity.

\subsubsection{Shape}\label{sssec:shape}
Similarly to rotation and size, the shape of an asteroid can be resolved, but only for large and nearby objects. Similar to the rotational state, the shape can also be estimated with time-domain analysis of a light curve, though with added complexity. Light curves are simplest to measure with optical photometry given their ability to gather enough SNR in short exposures compared with other wavelengths. Since optical photometry measures the brightness of reflected sunlight, the observed flux is proportional to area of the reflective surface apparent to the observer during the exposure time\footnote{This also changes with the Sun--asteroid--Earth angle, but much more slowly}. The relative flux observed over a period of rotation allows the shape to be constrained via the amplitude of the variations. Since asteroids often have complex shapes, tri-axial convex spheroid reflection models are typically assumed. A method to constrain the shape and rotational properties of an asteroid by inverting the light curve was developed by \citet{1992AA...259..333K, 1992AA...259..318K} and \citet{1993PhDT.......315L}. Optimization techniques advanced the method \citep{2001Icar..153...24K, 2001Icar..153...37K}, and the model was incorporated into a Bayesian statistical framework more recently with ground and space-based data of varying sparsity \citep{2020AA...642A.138M}.

\section{NEO Detection Surveys}\label{sec:detection}
In the previous section, I discussed the various NEO missions that flow from the text of the NEO Survey Act. In this section, I home in on the detection mission and discuss the major surveys of the present and future. I will assess their abilities to address the characterization and tracking missions as well. 

\subsection{Present Surveys and Their Limitations}
Today, two ground-based surveys discover over 75\% of all the NEOs. These are the Catalina Sky Survey \citep[hereafter CSS;][]{1998BAAS...30.1037L} and the Panoramic Survey Telescope and Rapid Response System \citep[hereafter Pan-STARRS;][]{2002SPIE.4836..154K}. CSS has three telescopes, two with wide field of view cameras optimized for survey-based discovery and one that targets NEOs for IOD refinement. CSS has detected around 30\% of the large NEOs over the past few years. Pan-STARRS has two telescopes with wide field of view cameras optimized for survey-based discovery. Pan-STARRS has detected around 55\% of the large NEOs over the past few years. Both CSS and Pan-STARRS operate from the Northern Hemisphere, which limits the overall NEO detection rate since the ecliptic plane extends into the southern sky. The remainder of 140 meter and larger NEOs are discovered by a handful of asteroid and synoptic surveys. The largest contributors are the Asteroid Terrestrial-impact Last Alert System \citep[or ATLAS;][]{2018PASP..130f4505T}, the Zwicky Transient Facility \citep[ZTF;][]{2019PASP..131a8002B}, and NEOWISE {\citep{2011ApJ...731...53M}}, which each detect a few percent of the large NEOs. ATLAS has a smaller aperture and is better at nearby NEOs given its rapid cadence. ZTF is not a dedicated NEO survey, but given its 47 square degree field of view and high cadence, it frequently detects them. NEOWISE, is an extended NEO-focused mission using the Wide-field Infrared Survey Explorer \citep[WISE;][]{2010AJ....140.1868W}, which was an all-sky survey in four MIR channels before its mission ended. The telescope was revived for NEOWISE with just the two shorter IR channels since the two longer wavelength channels ran out of refrigeration. 

\begin{figure}[htb!]
  \centering
  \includegraphics[width=\textwidth]{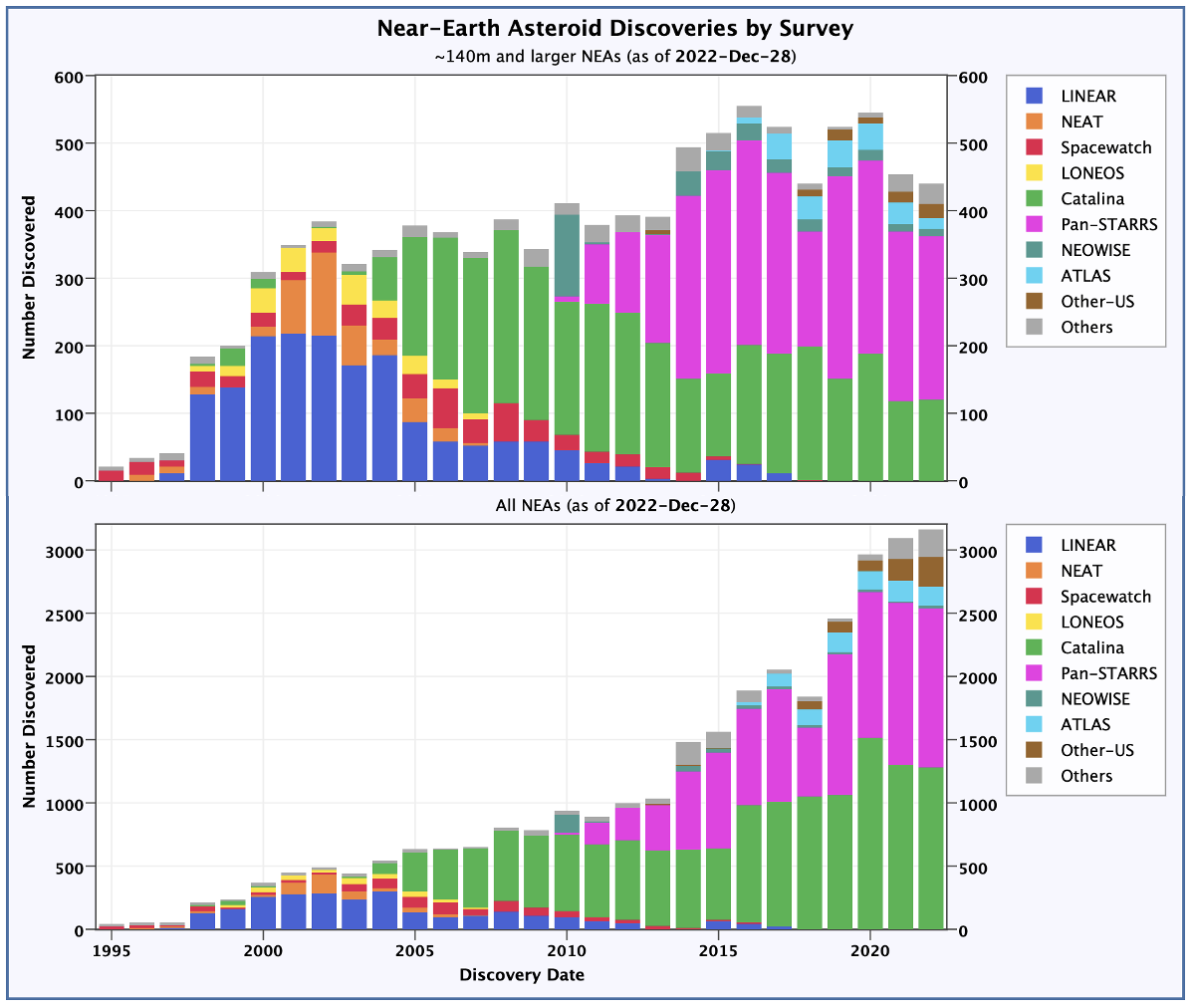}
  \caption{Detection statistics of 140 meter and larger NEOs (top panel) and NEOs of all sizes (bottom panel) by survey and year through 2022-Dec-28. The continued increase in detections of NEOs of all sizes suggests increased survey capabilities; however, the slower increase in larger NEO discoveries indicates that the pool of remaining large NEOs is becoming harder to detect. Images courtesy of the Center for Near-Earth Object Studies NASA/JPL-Caltech available at \url{https://cneos.jpl.nasa.gov/stats/}.}
  \label{fig:5}
\end{figure}

\begin{figure}[htb!]
  \centering
  \includegraphics[width=\textwidth]{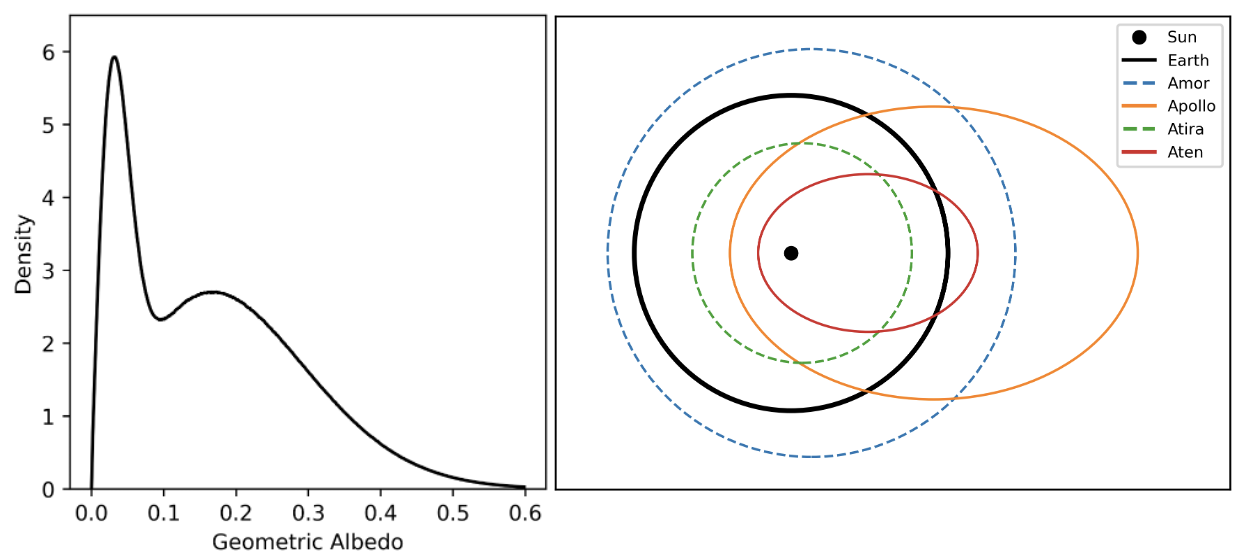}
  \caption{\emph{Left:} Albedo distribution for NEOs derived from WISE \citep{2016AJ....152...79W}. \emph{Right:} Example orbits for the four NEO types in relation to Earth and Mars. Apollos and Atens (solid ellipses) are Earth-crossing, whereas Amors and Atiras (dashed ellipses) are not. PHAs are those that cross within 0.05 AU and are larger than 140 meters.}
  \label{fig:6}
\end{figure}

At the present rate, it would take 40 years to meet the 90\% mandate without accounting for the slowing detection rate as the pool of undetected asteroids becomes saturated with those that are undetectable with the current surveys. This is evident by comparing the increasing rate of discovery in the bottom panel of Figure \ref{fig:5} versus the top panel. There are two primary drivers making the remaining population more difficult to detect over time. First is the albedo, which is the fraction of incident light reflected by the surface of the asteroid. The left panel of Figure \ref{fig:6} shows the albedo distribution of NEOs derived with WISE data \citep{2016AJ....152...79W}. There are numerous with albedo of only a few percent. Since most NEOs have been detected with reflected light, undetected NEOs are likely over-represented by those with low albedo. This over-representation will grow with time. The second difficulty is that many NEO orbits are inherently challenging to observe, especially with ground-based surveys, which contribute the vast majority of discoveries. The orbit of a given NEO will dictate the distances and reflection angle, which, along with the albedo, govern the apparent brightness of the asteroid for the observer.

NEOs can be split into four groups depending on just two of their orbital parameters. The groups are named after the original archetypes (1221 Amor, 1862 Apollo, 2062 Aten, and 163693 Atira). According to CNEOS, as of the end of 2022, there are 11,110 known Amor asteroids, 17,451 known Apollo asteroids, 2427 know Aten asteroids, and 28 known Atira asteroids. Apollos and Amors have semi-major axes larger than Earth. Apollos have large enough eccentricities such that they are Earth crossing, or put another way, their perihelia are smaller than Earth's aphelion ($q < 1.017$ AU). Meanwhile, since Amors perihelia are larger than Earth's aphelion ($q > 1.017$ AU), they are not Earth-crossing. Atens and Atiras are analogous with semi-major axes smaller than Earth, and Atens are the Earth crossing sub-population. See the right panel of Figure \ref{fig:6} for a schematic of each of these types of NEOs. Potentially hazardous asteroids (PHAs) are Apollos and Atens larger than 140 meters that cross within 0.05 AU of Earth’s orbit. Most NEOs originate from ejected MBAs, so it is not surprising Amors and Apollos are the most populous types given their proximity to the MBA source. They are also the easiest to detect with ground-based surveys since they spend most (or all for Amors) of their time outside of Earth's orbit. Atens and Atiras, on the other hand, are under-represented in the detected population because they are more difficult to observe from the ground. Simulations also suggest Atens and Atiras are intrinsically rarer types of NEOs \citep{2018Icar..312..181G}.

Beyond these orbital groupings, the other orbital parameters can make some asteroids evade detection for long periods of time. For example, a very Earth-like orbit exactly out of phase from Earth would be positioned behind the Sun for many years. On the timescale of the NEO surveys to date, such an asteroid could be undetectable with night-time surveying from the ground. As another example, an asteroid with a large inclination could evade detection from many asteroid surveys since they preferentially observe the ecliptic plane where asteroids are most numerous. Furthermore, for all except NEOWISE, the vast majority of NEOs are discovered via ground-based surveys, and most of these are carried out in the northern hemisphere. Ground-based optical surveys are unable to observe on the Sun side of the ecliptic plane (with the exception of twilight observing or high airmass observing just after evening twilight and before morning twilight; however, these observing modes are sub-optimal given the variable bright sky background and higher atmospheric extinction. Indeed, the majority of NEOs discovered today are close to Solar opposition.

\subsection{Planned NEO Surveys}
While the planned NEO surveys discussed in this section are primarily for detection, I will also discuss their contributions to the detection, tracking, and characterization. I will continue to return to this breakdown throughout this paper to highlight capabilities and gaps of the various surveys and follow-up instruments over the next two sections.

\subsubsection{NEO Detection Mission}
The observational challenges that the current NEO surveys are plagued by helps explain the designs of the next generation of asteroid detection missions. There are two primary surveys that will satisfy the NEO detection mission. The first is Rubin/LSST, which will have a 3.2 gigapixel camera that will cover the entire available night sky every few nights for a decade to extremely faint detection limits. Median single-epoch $5\Sigma$ detection limits are $u=23.9$, $g=25.0$, $r=24.7$, $i=24.0$, $z=23.3$, and $y=22.1$ in a 30 second exposure. First light is expected in 2024, and although the survey is not optimized for asteroid detection, the leap forward in sensitivity compared with ZTF (which carries out a similar survey) means a large fraction of the remaining NEOs will be discovered with Rubin/LSST alone. 

Several studies estimated that if the other surveys continue running at the current capabilities, $\sim$75\% of the 140m and larger NEOs will be discovered after a decade of Rubin/LSST \citep{2007IAUS..236..353I,2016AJ....151..172G,2017arXiv170506209C,2018Icar..303..181J}. Rubin/LSST will solve a few of the detection challenges described above. First, its increased sensitivity and wide area coverage will enable fainter asteroid detections over a broader portion of orbital space, which will probe lower albedo NEOs as well as NEOs with sub-optimal viewing geometry for ground-based observing. This will especially be true for the proposed twilight survey \citep{2018arXiv181200466S}, which will observe lower Solar elongations than CSS and Pan-STARRS do with much higher sensitivity than other instruments that carry out such surveys such as ZTF \citep{2020AJ....159...70Y}. Second, additional survey time with Rubin/LSST has been proposed to finish the 90\% detection threshold; \citet{2018Icar..303..181J} estimated that two additional years would finish the task, which is explained by the point raised above where out-of-phase asteroids will have additional time to get into phase with the ground-based observing scene. However, this proposal is an expensive one. The estimated yearly operating budget for LSST operations is nearly \$100M. 

\begin{figure}[htb!]
  \centering
  \includegraphics[width=\textwidth]{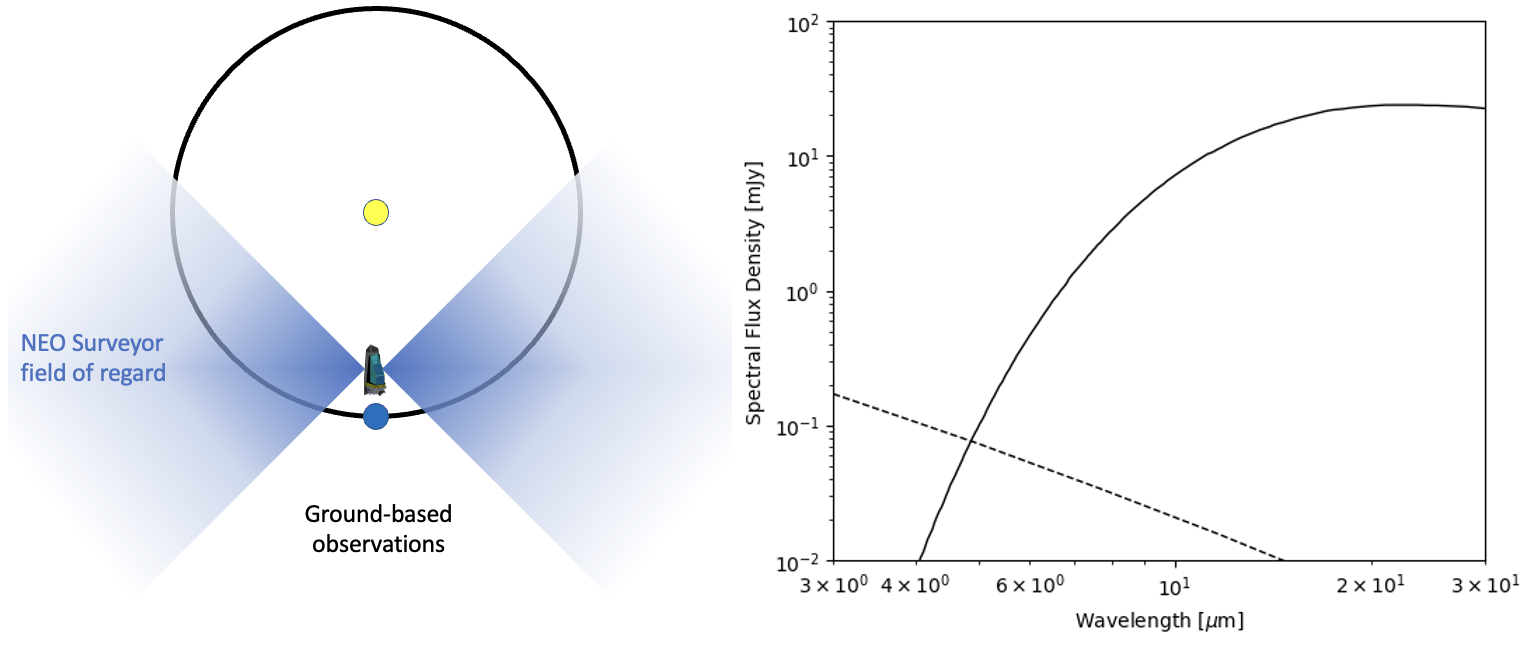}
  \caption{\emph{Left:} Schematic showing the NEO Surveyor field of regard, which is complimentary to ground-based surveys, which prioritize zenith observations during the night and thus most easily observe the white sector to the bottom of the figure. Rubin/LSST has a twilight survey planned, which will cover much of NEO Surveyor's field of regard, but with diminished capabilities due to the bright and variable twilight sky background. \emph{Right:} Reflected (dashed) and thermal (solid) contributions to the IR emission of NEOs assuming a large NEO with typical albedo and emissivity under the NEATM \citep{1998Icar..131..291H}.}
  \label{fig:7}
\end{figure}

The second major survey NASA’s proposed NEO Surveyor \citep{2022DPS....5440902M}, which is a space-based MIR system, which will operate at Earth-Sun L1 in a halo orbit \citep[see][for the original concept of a halo orbit around Lagrange points]{Farquhar1966StationkeepingIT}. This position inside of Earth’s orbit will allow for optimal viewing of the regions that are inherently challenging to observe from Earth (see left panel of Figure \ref{fig:7}). Furthermore, the MIR sensor will detect thermal emission of NEOs rather than reflected light, so it is less susceptible sensitivity losses from the low angle of reflection and low albedos. NEO Surveyor will launch no earlier than 2028, but together with the ground-based surveys it will enable the completion of the NEO detection mission by the early 2030s. Acting by itself, NEO Surveyor will detect about two thirds of the target NEOs in five years or 90\% in twelve years \citep{2022DPS....5440902M}. The survey has limited lifetime given its halo orbit, which requires constant station keeping. In the right panel of Figure \ref{fig:7}, an example thermal and reflected emission spectrum is shown under the near-Earth asteroid thermal model \citep[NEATM;][]{1998Icar..131..291H}.

\subsubsection{NEO Tracking Mission}
The observing cadence for Rubin/LSST is designed such that NEOs will be self-recovered a few times per lunation such that a discovery and IOD may be reported to the MPC \citep{2018Icar..303..181J, 2019ApJ...873..111I}. Rubin/LSST will implement the Moving Object Processing System (MOPS) originally developed for Pan-STARRS for track linking \citep{2013PASP..125..357D}. Typical NEOs will be rediscovered a couple hundred times over the 10-year survey such that orbits may be refined and updated with Rubin/LSST alone; however, the gap between the discovery exposures and new recovery exposures will often be large. Thus, additional follow-up instruments will be important for refining the trajectory. However, Rubin/LSST will require very sensitive follow up observations given its faint detection limits. This is especially the case for nearby NEOs that are intrinsically small and quickly move away from Earth to lower Solar elongation.

NEO Surveyor is still in the planning and mission design phase of its life cycle, but it has a similar operational plan as Rubin/LSST in that it is prioritizing survey-based discovery, and IODs will be issued based upon self-recovery over a short period of time. However, there is a need for higher-fidelity IODs given that follow-up instruments on the ground will not be able to observe newly discovered NEOs until they move to larger Solar elongations (see left panel of Figure \ref{fig:7}).

\subsubsection{NEO Characterization Mission}
Here I will discuss the ability of Rubin/LSST and NEO Surveyor to satisfy the NEO characterization mission. Both surveys will have new capabilities beyond that of the current state-of-the-art surveys; however, there are some capability gaps, which I will discuss.

Rubin/LSST is an optical and NIR photometry instrument with six broad-band filters covering the visible to NIR wavelengths. Its unique capability will be color determinations across between 350 nm and 1050 nm. LSST's cadence design cycles through its six filters and also returns to recover NEO discoveries a few times per lunation, so it is likely that a discovery and IOD will be made with multiple filters. As I discussed in \S\ref{sssec:material}, these color-indices are useful for constraining the type of asteroid. In Figure \ref{fig:3}, I showed the large spread in color indices and broad-band magnitudes for typical asteroids from all 24 asteroid types derived via visible to NIR spectroscopy. Given the photometric precision of Rubin/LSST, the colors measured with just the few observations needed to discover and establish an IOD will be sufficient for rough asteroid characterization. Colors that span larger wavelength baselines such as $g-y$ and $r-y$ are the most constraining. However, many NEOs rotate many times per day, and the colors will be confounded with potential periodicity in the light curve, which will not be measured enough to correct, so color indices could be off by as much as the peak-to-trough difference in magnitude. Thus, the Rubin/LSST derived colors estimated during detection are at best only useful for roughly constraining material properties of a newly discovered NEO. They will become much more accurate after the rotational state is determined and the discovery measurements can be phase-wrapped. On longer time scales, Rubin/LSST will detect the same NEO tens to hundreds of times. This could potentially allow for time-series analysis of the light curve; however, this will require extremely accurate phase wrapping, which is likely sub-optimal compared to more timely follow up from dedicated instruments. An exception to this is possible for NEOs that track through a deep-drilling field, which may have the requisite cadence to enable light curve analysis. 

\begin{figure}[h!]
  \centering
  \includegraphics[width=0.66\textwidth]{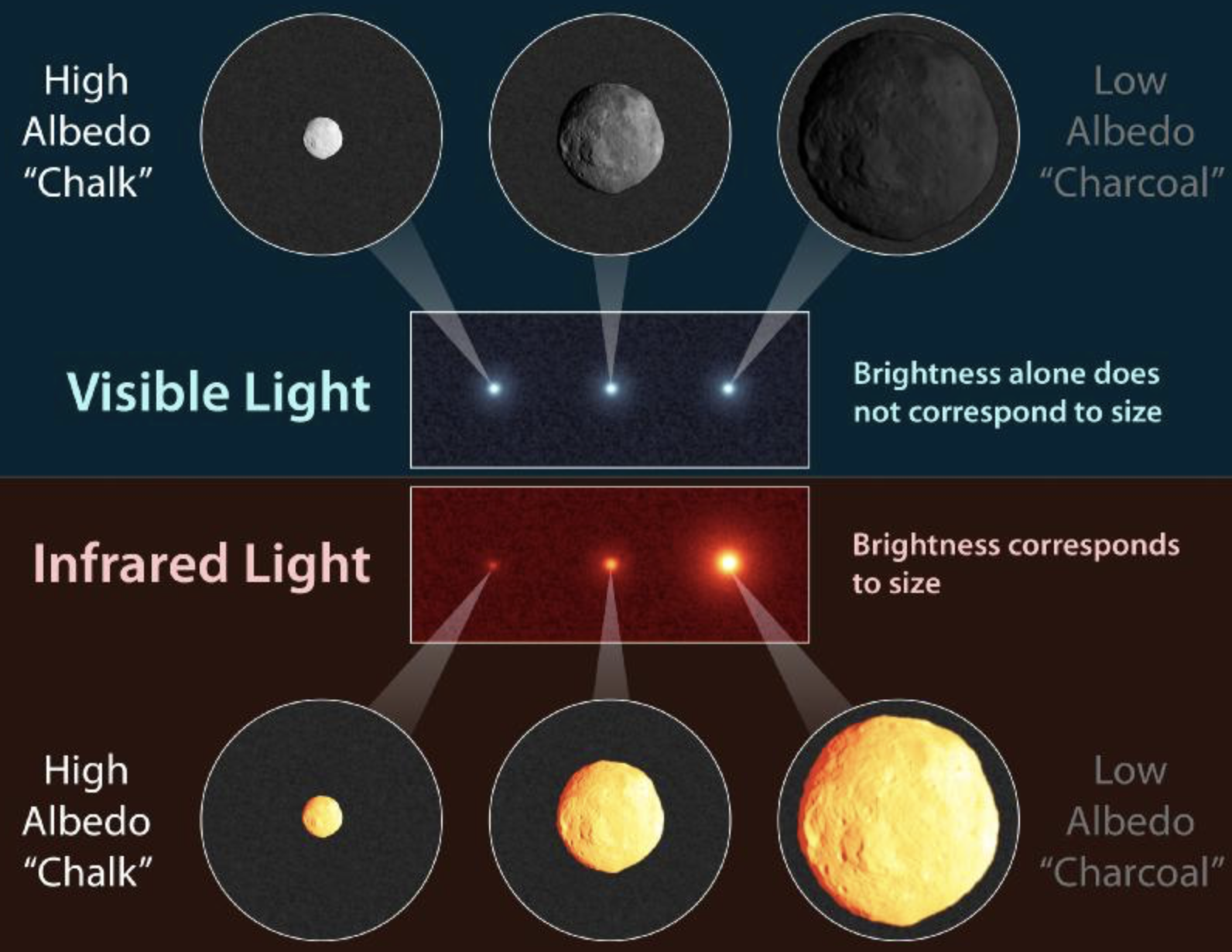}
  \caption{Schematic showing the value of combining visible and MIR (thermal) observations of NEOs. With visible light alone, there is a degeneracy between albedo and size. With MIR alone, only the mass is determined but not the albedo. By combining the two, the mass is determined with the MIR and the albedo is determined with the combination. If there is also an estimate of the material properties than the size and thus density may also be determined. Image credit: NASA/JPL-Caltech available at \url{https://www.nasa.gov/mission_pages/WISE/multimedia/gallery/neowise/pia14733.html}.}
  \label{fig:8}
\end{figure}

NEO Surveyor is a MIR imager with two pass-bands. Beyond about 4.5 $\mu$m, the thermal emission dominates the reflected sunlight (see right panel of Figure \ref{fig:7}). NEO Surveyor's instrument will effectively measure the heat signature of a NEO, which means the flux will be proportional to the size of the asteroid. This is unique to MIR sensors, and is a valuable piece of information for planetary defense. An important follow-up for any NEO Surveyor discovery will be an optical flux estimate. When combined with the thermal IR measurement that gives the size of the asteroid, the albedo can be deduced (see Figure \ref{fig:8}). For NEO Surveyor acting alone, this will be useful for when known objects (predominantly discovered via optical photometry) are recovered. However, given the orbital regime that NEO Surveyor will observe, new discoveries will need timely follow up with optical photometry to quickly determine these additional properties. As discussed above, this is challenging for ground-based follow up instruments. NEO Surveyor will not be as useful for shape or rotational property estimation since the MIR flux is not as strongly dependent on phase as optical photometry. NEO Surveyor's two MIR pass bands will give some indication of potential emissivity variation of asteroids as they rotate, which could be used to constrain the material properties, but it will not be as strong as multiple broad-band visible to NIR photometry \citep{2009Icar..202..160D}. 

\subsection{Gaps in the NEO Mission Space}
\begin{figure}[!ht]
  \centering
  \includegraphics[width=\textwidth]{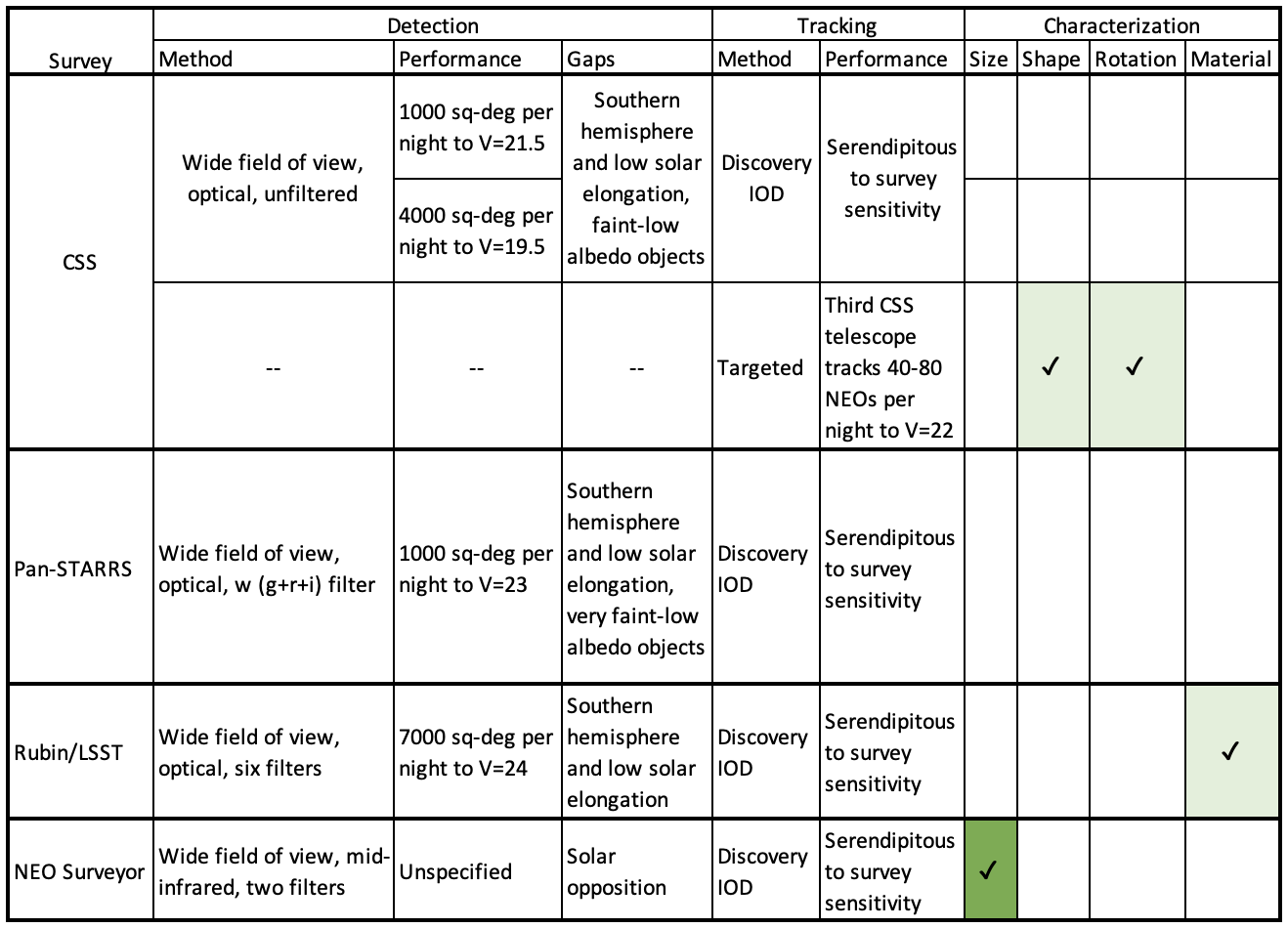}
  \caption{Breakdown of the two major present and two major future NEO surveys. For the characterization mission check marks are placed in cells where the survey plausibly contributes to the specified feature within normal survey operations in a timely manner. The level of shading indicates the relative performance of the survey compared with a dedicated follow up instrument. CSS and Pan-STARRS are largely similar except for the third CSS telescope, which offers some time-domain characterization in addition to its purpose of refining NEO tracks. Rubin/LSST uniquely offers multi-band photometry enabling material constraints based on color information. NEO Surveyor independently constrains the size of an asteroid..}
  \label{fig:9}
\end{figure}

Previously in this section, I discussed the major NEO surveys of the present and future, and I broke down their contributions to the detection, tracking, and characterization missions of the NEO Survey Act. Here I summarize the discussion to highlight the components of the various missions that remain for follow up facilities to tackle. This information is summarized in Figure \ref{fig:9}. 

No single survey is capable of detecting all NEOs. Ground based surveys are limited by their aperture size; though, Rubin/LSST will be sensitive enough to detect far more NEOs that enter its field of view that CSS or Pan-STARRS. They are also limited by their inability to observe during the daytime, and so low Solar elongations are unobserved. Conversely, NEO Surveyor can observe lower Solar elongations, but it cannot observe at Solar opposition due to the Earth. 

Only CSS is capable of refining the track of NEOs; though, its capability is niche because it has a third telescope dedicated to this task. However, the throughput of a single ground-based sensor is limited to fewer than 100 NEOs per night. While this likely satisfies refining the orbits of discoveries today, future surveys will detect far more NEOs that are too numerous to follow up with a single telescope or are too faint to follow up from 1 meter class optical telescopes. 

Once both Rubin/LSST and NEO Surveyor are operational, asteroids observed by both surveys will have a good track, size, and rough material property constraints. However, this characterization will not be timely since NEO Surveyor's and Rubin/LSST's detections will be temporally separated due to their different survey regimes. Furthermore, much of the characterization from these two surveys will be serendipitous in nature; that is, the surveys are designed for rapid recovery of a few exposures in order to issue a discovery and IOD, but detections that come afterward are serendipitous and may come months or years later. This is most especially true for the time-domain analysis required to estimate shape and rotational properties. Neither Rubin/LSST nor NEO Surveyor will provide high fidelity estimates of the shape or rotational properties. 

\section{NEO Follow Up}\label{sec:followup}
Today, after discovery and IODs are reported by the discovery survey (typically CSS or Pan-STARRS), they are posted to the MPC, which maintains a list of NEOs in need of further observation\footnote{\url{https://minorplanetcenter.net/iau/NEO/toconfirm_tabular.html}}. The tracking mission is often carried out by amateur astronomers around the world, who observe and report astrometry and photometry back to the MPC, which recomputes orbital states frequently and moves confirmed NEOs off this list. In this process, most large NEOs are observed rapidly for a few days after discovery. In principle, this could enable time-domain analyses for shape and rotational properties; however, the MPC does not have strict requirements on data quality. MPC photometry precision is often unreported and assumed accurate only to 0.1 magnitudes. Many amateur astronomers do not calibrate their observations to the level that is done at professional facilities. Furthermore, their observations typically come from sites with poor atmospheric conditions with poor seeing, high levels of aerosols from air pollution, and bright sky backgrounds from scattered terrestrial light; thus, it is challenging to estimate uncertainties properly for the relevant time-domain analysis and color index estimate. 

For color index estimation, two magnitude estimates with an uncertainty of 0.1 mag from the MPC database leads to an uncertainty in the estimated color index by of $\sqrt{2\times0.1^2}\approx.14$ magnitudes. Most amateur astronomers observe with visible sensors and filters, so such an uncertainty is comparable to the spread of colors for asteroid spectral classes presented in the bottom panel of Figure \ref{fig:3}; thus, typical MPC photometry is insufficient for color index estimation to the level of precision needed for constraining the material properties of NEOs. Nevertheless, the light curve and color information posted to the MPC has been collected by the Asteroid Lightcurve Database \citep{2009Icar..202..134W}. 

In the following subsections, I will review the major facilities that astronomers use for the NEO characterization mission. This will not be a comprehensive list. I will prioritize systems that are most useful for rapid characterization, and I will organize the facilities by their wavelength and observing modality and mention the various asteroid features that are probed by each facility. 

\subsection{Optical NEO Follow Up Facilities}

Optical photometry is the cheapest type of observation to acquire for the NEO characterization mission. The relatively low-cost for ground-based optical telescope time makes them particularly useful for building light curves for shape and rotational analysis. A system particularly well-suited for this purpose is Las Cumbres Observatory Global Telescope Network \citep[LCOGT;][]{2013PASP..125.1031B}, which is composed of twenty seven telescopes (ten 0.4 meter telescopes, fifteen 1 meter telescopes, and two 2 meter telescopes) at seven sites around the world. The telescopes have a variety of instruments including low and high-resolution optical spectrographs as well as photometric cameras with moderate field of view ($\sim\,$0.2 sq-deg) and full broadband filter sets. Each telescope is fully robotic, operates from optical to NIR wavelengths, and observations are scheduled in an optimized fashion across all observers to minimize slew time while taking into consideration the timeliness needed to achieve the observer's science goals. LCOGT serves planetary science well and has even dedicated key projects\footnote{\url{https://lco.global/science/keyprojects/}} to asteroid science. This key project has in part developed NEOExchange\footnote{\url{https://lco.global/neoexchange}}, which helps collect and easily queue observations with LCOGT \citep{2021Icar..36414387L}. LCOGT also participated in the global observations of the DART target Dimorphos \citep{2022LPICo2678.2534L}. 

LCOGT has recently partnered with the National Science Foundation to include NOIRLab's Community Science and Data Center, 4.1 meter SOAR telescope, and 8 meter Gemini-South telescope into a combined follow up network for high-profile transients detected with Rubin/LSST. This partnership is named the Astrophysical Events Observatories Network \citep[AEON][]{2020SPIE11449E..25S}. However, it remains to be seen how the NOIRLab facilities, which operate in a traditional model with blocks of consecutive nights that requiring the investigators to operate the instrument. Some telescopes have begun offering target of opportunity (ToO) time, where proposers have the ability to borrow time from traditional proposals if a transient phenomenon occurs. This methodology became popular after the detection of gravitational waves and full-spectrum electromagnetic counterparts \citep{2017ApJ...848L..12A}. While AEON will certainly be good for transient astronomy, NEOs make up only a portion of alerts delivered by LSST, which will also detect supernovae, all types of variable stars, minor planets throughout the Solar System, among others totaling $\sim\,$ 10 million alerts per night \citep{2019ApJ...873..111I}. A drawback for LCOGT and AEON is that they are entirely ground-based, so it will be unable to follow up most NEO Surveyor discoveries. 

Other groups have organized around NEO characterization. One notable group is the Mission Accessible Near-Earth Objects Survey \citep{2015LPICo1829.6038M}. MANOS has prioritized smaller NEOs and targeted them with 4--8 meter class telescopes both spectroscopically and photometrically \citep{2016AJ....152..163T, 2019AJ....158..196D}. To date they have reported 500 visible spectra and an optical lightcurve for over 700 NEOs\footnote{\url{https://manos.lowell.edu/observations/summary}}.

Optical cameras generally are very common on telescopes of all apertures including some of the largest telescopes worldwide. Modern sensors have also skewed toward wide field of view such as the Dark Energy Camera \citep{2015AJ....150..150F} and Hyper Suprime Cam \citep{2018PASJ...70S...1M}, both of which were designed to carry out wide-field surveys for cosmology but have since transitioned to general observing through their respective telescope proposal processes. Large aperture, wide field of view systems are particularly useful for recovering faint asteroids with uncertain orbital states such as faint NEOs detected nearby Earth which become too faint to recover with the NEO surveys. Such instruments are typically over-subscribed via their calls for observing proposals, and the typical minimum unit of time is a half or whole night; thus, singular observations to recover an asteroid are a challenging modality.

\begin{figure}[hbt!]
  \centering
  \includegraphics[width=\textwidth]{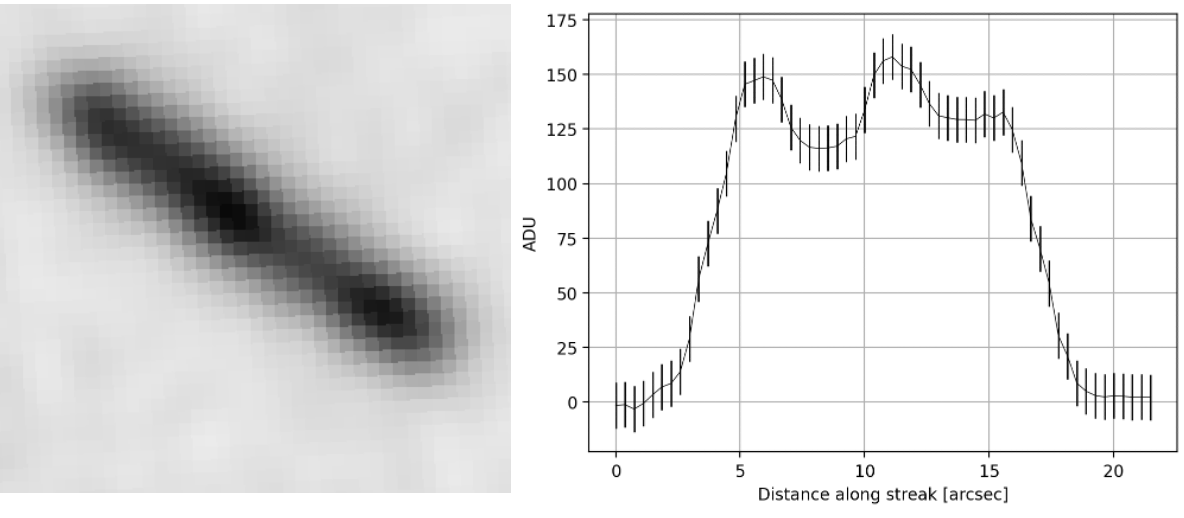}
  \caption{Observation of rotation within a single 5-minute exposure of 2010 XC$_{15}$ with the Nickel 1 meter telescope at Lick Observatory. Observations were taken by Connor Dickinson and Srujan Dandu of UC Santa Cruz under Nickel proposal 2022B-N006 (PI: Kirsty Taggart). \emph{Left:} 300-s B-band photometric image in reverse color. Flux variations along the streak are apparent. North is up and East is left. The proper motion was to the Northeast. The camera was readout with 2$\times$2 binning such that the pixels are 0.37$''$. The image was convolved with a symmetric two dimensional Gaussian kernel of width equal to the atmospheric seeing at the time of the exposure (FWHM$=1.29''$). \emph{Right:} Boxcar projection along the length of the streak. The width of the boxcar is $2\times$FWHM. The error bars represent the standard deviation in the pixel ADU values in the vicinity of the asteroid indicating that the flux variations are statistically significant.}
  \label{fig:10}
\end{figure}

Asteroid observations have been carried out on numerous smaller aperture optical telescopes with good success; for example, the 0.9 meter WIYN telescope was used to precisely detect and estimate orbital parameters of asteroids to faint magnitudes using a stacking technique known as ``shift and stack'' or ``digital tracking'' \citep{2015AJ....150..125H}. Recently, I used the 1 meter Nickel telescope to obtain multi-band photometry covering the visible wavelengths to estimate asteroid classifications of NEO that came within a few lunar distances in late 2022 (Golovich et al. in prep). One of the targeted asteroids exhibited fast rotation apparent in a single five minute exposure (see Figure \ref{fig:10}). 

Optical spectroscopy is also available at many observatories. Spectroscopy is inherently more constraining since a spectrograph gives a data-point for model comparison per spectral resolution unit. Even a low-resolution spectrum such as those offered by LCOGT's offers far more parameters for model comparison; however, the trade is for exposure time. Using the LCOGT's exposure planning tools, 21 minutes (including overhead) are needed to observe the spectrum of a NEO at g=16 at a spectral resolution of R$=500$ to a SNR of 10 per resolving unit; whereas, the same asteroid could be observed eight times in six filters spanning the same wavelength range, which would offer a coarse replication of the SED as well as dozens of observations spanning 20 minutes for time-series analysis for shape and rotation. 

\subsection{Infrared NEO Follow Up Facilities}
IR wavelengths range from $\sim$ 0.75 to 1000 $\mu$m, but can be further split into NIR (0.75-3 $\mu$m), MIR (3-25 $\mu$m), and far-IR (25-1000 $\mu$m). Asteroids are typically studied with NIR and MIR observations with NIR dominated by reflected light and MIR dominated by thermal emission above 3--4 $\mu$m (see right panel of Figure \ref{fig:7}). 

IR astronomy has a long history due to the gaps in the atmospheric absorption spectrum. The gaps are given names, which follow through to the names of the photometric filters. The most common NIR bands are I ($\sim0.8\,\mu$m), Z ($\sim0.9\,\mu$m), Y ($\sim1.0\,\mu$m), J ($\sim1.2\,\mu$m), H ($\sim\,1.6\mu$m), and K ($\sim\,2.2\mu$m). I, Z, and Y-bands are commonly coupled with optical astronomy facilities. J, H, and K-bands were the filters used for the Two-Micron All-Sky Survey \citep[2MASS;][]{2006AJ....131.1163S}, which provide the canonical calibration data set for nearly all NIR astronomical observations. 

There are several major ground-based IR facilities; however, few are open to NEO observations through general calls for proposals. The most NEO-focused instrument is NASA's Infrared Telescope Facility (IRTF), which has a guaranteed minimum of 50\% time toward planetary science. IRTF has both imagers and spectrographs covering NIR and MIR. As discussed above, NIR is predominantly reflected light, but it is highly constraining for material classification, especially when paired with optical photometry. This is evident from the slope of the SEDs in the top panel of Figure \ref{fig:3} and the large spread in colors between the various asteroid types, especially the g$-$NIR color index in the bottom panel of Figure \ref{fig:3}. IRTF has been used for much of the research discussed in this paper regarding remote-sensing enabled asteroid material studies and classifications. The MIT-Hawaii Near-Earth Object Spectroscopic Survey (MITHNEOS)\footnote{\url{http://smass.mit.edu/minus.html}} has had a long-running campaign of obtaining visible to NIR spectroscopy for large and nearby NEOs. Much of their data has been made public \citep{2009Icar..202..160D, 2014Icar..233....9P, 2019Icar..322...13D, 2019Icar..324...41B}. Over twenty years, MITHNEOS has measured $\sim$1600 NEO spectra spanning visible to NIR wavelengths. 

Many large-aperture telescopes around the world have IR sensors coupled with adaptive optics systems. While these enable resolved imagery of large and nearby NEOs, which give direct measurements of size, shape, and rotational properties, these exposures are expensive to carry out and time on 8$-$10 meter telescopes such as Subaru, Keck, and VLT are highly competitive.

The remaining notable IR instruments for NEO characterization are in space. The first is NEOWISE, which is a survey instrument and a precursor to NEO Surveyor. The second is James Webb Space Telescope \citep[JWST;][]{2006SSRv..123..485G}, which needs no introduction. JWST is certainly very powerful for NEO characterization. Its NIR and MIR spectrographs and imagers would enable extremely precise spectroscopy; however, it is perhaps overkill for most large NEOs. When JWST was being planned, white papers discussed NEO observations in terms of imaging NEOs as small as a meter \citep[e.g.,][]{2016PASP..128a8002T}. While there is value in studying small objects since most large asteroids are composite structures, it is clear that JWST is not going to meaningfully contribute to the NEO discovery, tracking, or characterization missions.

\subsection{Radar NEO Follow Up Facilities}
\begin{figure}[hbt!]
  \centering
  \includegraphics[width=0.66\textwidth]{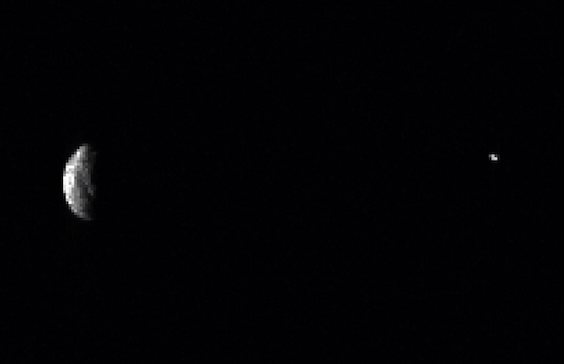}
  \caption{Goldstone Solar System Radar image of 2005 LW$_3$ showing a binary asteroid pair. The primary was measured at Image available at \url{https://echo.jpl.nasa.gov/asteroids/2005LW3/2005LW3.2022.goldstone.planning.html}. Reproduced here with permission, credit: NASA/JPL-Caltech.}
  \label{fig:11}
\end{figure}

Radar systems use a radio wave source and detect the reflected signal. Radar observations of asteroids are among the most useful for the NEO characterization mission, as I discussed above. Because they are often mounted on large radio dishes, they offer resolved imagery and can measure size, shape, and rotational properties. However, radar sensitivity falls as $r^{-4}$, so it is only viable for very nearby or very large NEOs. Nevertheless, when NEOs fly near Earth, radar observations should be prioritized. A recent NEO flyby of 2005 LW$_3$ resulted in an unexpected discovery. The asteroid's albedo was unknown prior to the close approach, so a nominal value of 0.15 was assumed, which suggested an object a $\sim100-200$ meters in diameter. However, observations with NASA's Goldstone Solar System Radar resolved a binary asteroid system\footnote{The discovery was reported in CBAT Electronic Telegram No. 5198 by S. P. Naidu, L. A. M. Benner, M. Brozovic, and J. D. Giorgini, of Jet Propulsion Laboratory.} with a primary diameter of 400 meters and a secondary diameter of 100 meters orbiting $\sim$4 km from the primary (see Figure \ref{fig:11}). The primary's albedo was estimated to be 0.02. 

Additional radar facilities could be opened in the future. The Green Bank Observatory has been used as a single dish receiver using the Goldstone Solar System Radar; however, there are plans to develop radar capabilities for it and the Very Large Baseline Array \citep{2022DPS....5421109P}. These advancements are needed to replace the lost capabilities since Arecibo's demise \citep{2022DPS....5451401V}. 

\section{Unique Capabilities of a NEO Characterization Constellation} \label{sec:constellation}
In \S\ref{sec:mandate}, I discussed the NEO Survey Act and various missions that follow from the language of the bill. In \S\ref{sec:detection}, I detailed the major detection surveys operating today and the two major surveys that will operate over the 2020s and 2030s. By the end of that time frame, it is very likely that 90\% of the 140 meter and larger NEOs will have been discovered, and (especially with Rubin/LSST) many more NEOs as small as a few meters will be discovered flying close to Earth. NEO Surveyor, on the other hand, will detect NEOs in portions of the Solar System that are difficult to observe from the ground. In Figure \ref{fig:9}, I detailed the gaps in capability of the four major surveys for the detection, tracking, and characterization missions. While Rubin/LSST and NEO Surveyor will offer more characterization than previous surveys, they will not solely (or together) meet the NEO Survey Act's characterization mission. Thus, in \S\ref{sec:followup}, I detailed the methods and instruments astronomers use to follow up newly discovered NEOs. One key takeaway is that the networks of follow-up instruments are entirely on the ground and are largely composed of only optical sensors. Another takeaway is that there are no sensors dedicated to tipping and queuing newly discovered asteroids, and those that do exist (such as LCOGT) will be inundated with observing requests from multiple science cases. Asteroid alerts from Rubin/LSST will be only a portion of the millions of transient alerts released nightly. Finally, there is an inability writ-large to follow up discoveries made by the NEO Surveyor with low Solar elongation. This is particularly problematic since optical photometry follow up of NEO Surveyor's discoveries will directly measure the albedo (see Figure \ref{fig:8}).

\begin{figure}[hbt!]
  \centering
  \includegraphics[width=\textwidth]{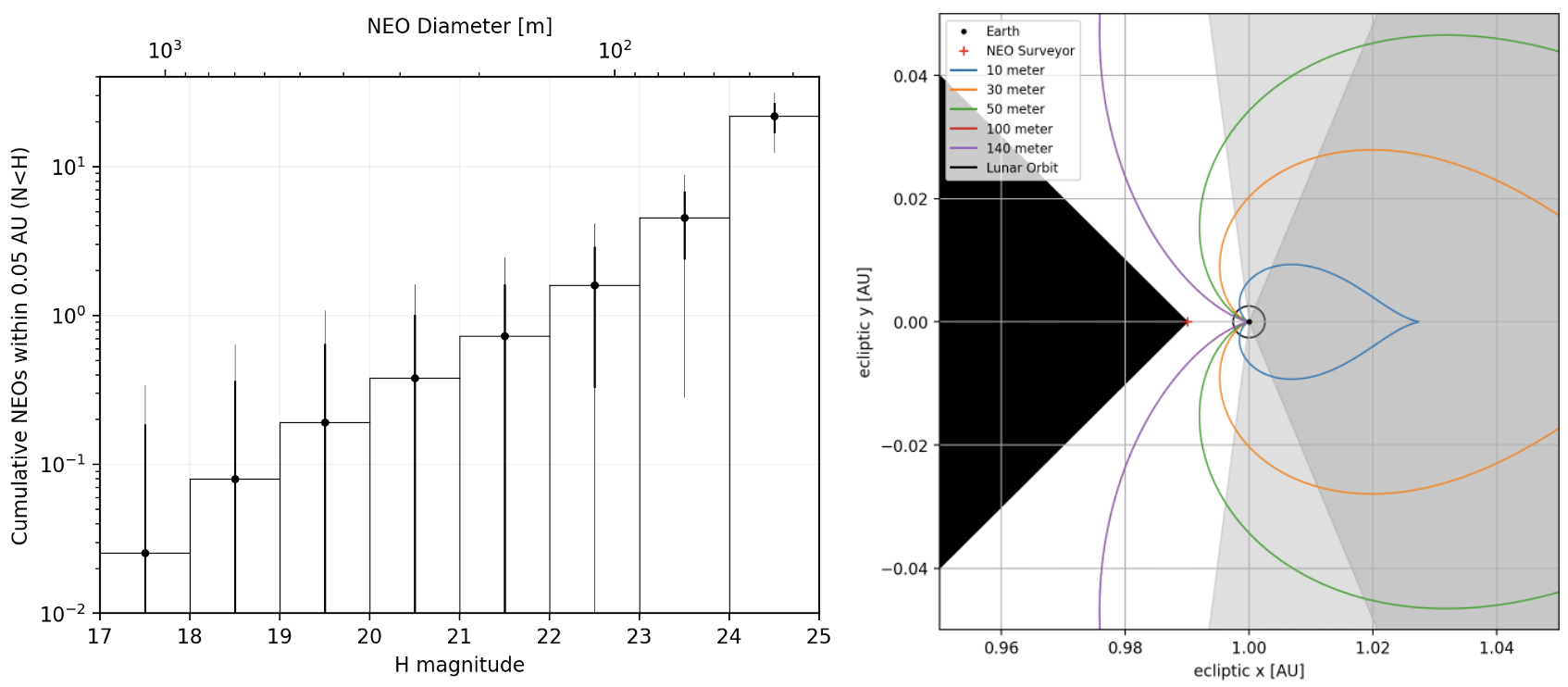}
  \caption{\emph{Left:} Monte Carlo simulation of the cumulative ($N(H)<H$) number of NEOs within 0.05 AU as a function of size assuming the simulated NEO orbital catalog of \citet{2018Icar..312..181G}. The black histogram and points indicate the mean of the Monte Carlo samples, and the thick and thin vertical lines indicate the 1$\sigma$ and 2$\sigma$ confidence intervals, respectively. The size axis at the top of the figure assumes an albedo of 0.15 and the size-magnitude relationship of \citet{1997Icar..126..450H}. \emph{Right:} Spatial reachability volumes for asteroids of various size. The black sector shows the Solar exclusion zone for the NEO Surveyor, which is positioned at Earth--Sun L1. The Moon's orbit is indicated for reference. The dark and light gray sectors show where 68\% and 95\% of Rubin/LSST pointings will occur, respectively. The white sectors show the regimes where ground-based observatories do not observe frequently. The Sun is located at $(x,y)=(0,0)$, which is relevant for the asteroid apparent magnitude contours.}
  \label{fig:12}
\end{figure}

There are an estimated 30,000 NEOs larger than 140 meters \citep{2012ApJ...752..110M, 2015Icar..257..302H, 2016Natur.530..303G, 2017Icar..284..114S}, today about 35\% have been discovered. As mentioned in \S\ref{subsec:size_thresh}, this is a somewhat arbitrary cutoff, especially given that there is enhanced risk from smaller (and much more numerous NEOs). \citet{2018Icar..312..181G} provides a simulated NEO catalog for objects as faint as H=25 (or 34 meters assuming 0.15 geometric albedo). To estimate the number within proximity to Earth at any given time, a Monte Carlo simulation was executed. For each sample, the 802,000 asteroid orbits from \citet{2018Icar..312..181G} and Earth were propagated on their orbits to a randomly selected time within the next 100 years of the orbital epoch (set to be 1 January 2023). The distance was computed to Earth and the $H$ magnitude was recorded for any asteroids within 0.05 AU. Orbits were instantiated along with Earth at random times within the next 100 years to provide for a good amount of mixing between the NEO orbital periods with respect to Earth. The left panel of Figure \ref{fig:12} shows the number of NEOs within 0.05 AU ($\sim$19.4 lunar distances) as a function of absolute magnitude and size (assuming an albedo of 0.15).  

The right panel of Figure \ref{fig:12} shows the spatial reachability volumes projected onto the ecliptic plane for various sized NEOs at $V=20$. The HG asteroid flux model \citep{1989LPI....20..375H, binzel1989asteroids} was assumed with G$=0.15$. The black sector is the Solar exclusion regime for the NEO Surveyor, which is positioned at Earth--Sun L1. Figure \ref{fig:12} demonstrates that a LEO constellation could characterize NEOs that are difficult to observe with ground-based assets (white sectors). The gray sectors show the regions that Rubin/LSST will observe. The choice of $V=20$ for the reachability volumes was made to demonstrate what would be feasible for a small optical telescope in space. Using an exposure time calculator assuming modern CMOS sensor characteristics (low dark current and read noise), and assuming the background was solely from the zodiacal light \citep{zodi}, it was estimated that 20$^th$ magnitude was obtainable within a few tens of seconds for optical filters and an unfiltered NIR camera. 

A constellation of small satellites in LEO equipped with visible and NIR sensors would be capable of filling the gaps left open by Rubin/LSST, NEO Surveyor, and ground-based follow up facilities like LCOGT. They could be tasked directly from the discovery lists or alert streams sent out by the surveys, and the increasing affordability of launching to LEO would enable the constellation to grow over time and keep pace with discoveries made by Rubin/LSST and NEO Surveyor. When there are too few new NEO discoveries to track at any given time, they could be tasked to further characterize known NEOs in the MPC. 

While the detection alert system for NEO Surveyor is not defined at this point (beside presumably to send discoveries to the MPC), Rubin/LSST will have alerts sent out within a minute of acquisition. A number of alert brokers have been proposed to curate the alert stream for individual science cases. One such broker, Automatic Learning for the Rapid Classification of Events (ALeRCE) will include asteroid classification labels, and has been tested on ZTF \citep{2021AJ....161..141S, 2021AJ....161..242F, 2021AJ....162..231C}. These curated alerts could be used to tip and queue the sensors from a central scheduler similar to the way LCOGT schedules their observations across many telescopes. 

With filters such as those presented in Figure \ref{fig:3}, a LEO constellation could identify asteroid classification with a series of colors and magnitudes measured to infer the underlying SED. Studies linking SED classifications for NEOs are linked to meteorite and sample return mission studies indicating known minerals \citep{2009Icar..202..160D, 2011Sci...333.1113N, 2012AGUFM.P53C..02P, 2022Icar..38014971D, 2022AJ....163..165M, 2023Icar..38915264D}. Furthermore, numerous telescopes could observe and hand-off targets to the next sensor as they orbit Earth in such a way that would build high fidelity and well-sampled light curves without gaps in coverage that are sufficient for shape and rotation inference, as I discussed in \S\ref{sssec:shape}. 

\begin{figure}[hbt!]
  \centering
  \includegraphics[width=\textwidth]{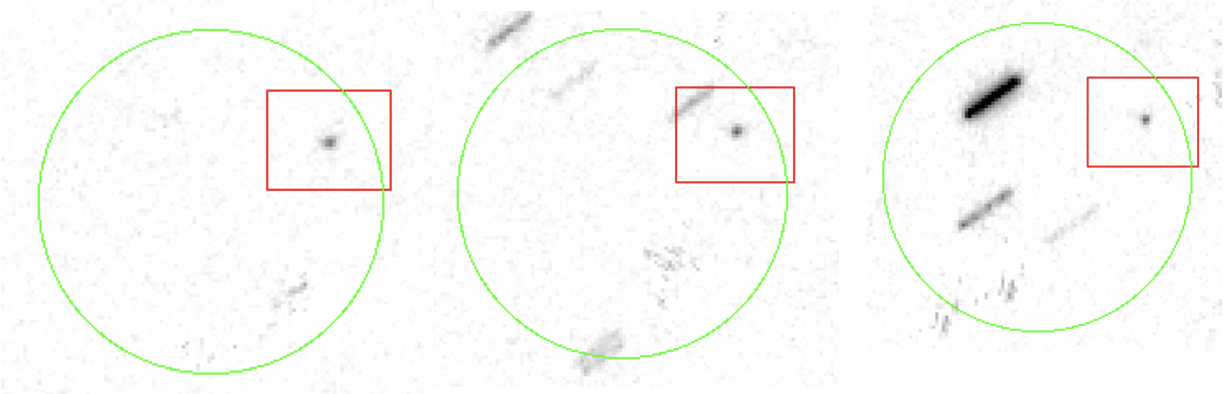}
  \caption{GEOStare2 images of 2005 LW$_3$ taken on 23 November 2022. Each image is a reprojected along the asteroid's relative motion with respect to the stars, which were tracked during a series of 1 second exposures. We were able to estimate a $\sim$200 minute rotational period, which was later confirmed by the Goldstone Radar images.}
  \label{fig:13}
\end{figure}

If the constellation was distributed along a few orbital inclinations in LEO, just a handful of sensors could essentially observe anywhere in the sky including at low Solar elongations. For example, I have recently tasked GEOStare2\footnote{GEOStare2 is a joint LLNL and Terran Orbital satellite used for space traffic management that we tasked to observe asteroids flying by Earth in late 2022. See  \url{https://www.llnl.gov/news/llnltyvak-space-telescope-goes-orbit} for a press release and outline of the mission.} to observe three NEOs flying by the Earth within a few lunar distances (Golovich et al. in prep). GEOStare2 is a pathfinder program run at LLNL for tracking space debris for space traffic management using a wide field of view 8.5 cm optical telescope. Coupling the space-based light curves with Lick multi-band optical photometry, we were able to characterize the asteroids rotational state and material properties. Figure \ref{fig:13} shows an example image taken by GEOStare2 of the asteroid 2005 LW$_3$. Though our GEOStare2 observations were not at particularly low Solar elongation, GEOStare2 is capable of observing within 50$^\circ$ of the Sun. Telescopes designed with simple baffling could observe closer, which would open the door to following up all NEO Surveyor detections in near real time. This capability would be unique to space-based follow up and tracking of newly discovered NEOs, so it is the only way to meet the tracking and characterization portions of the NEO Survey Act. Finally, the wide field of view of LSST, ATLAS, CSS, and Pan-STARRS makes it likely that small but nearby NEOs are discovered. These fly by Earth in as short as a few hours. The ability to characterize these NEOs would begin to unlock information about the composition of rubble pile NEOs, but it can only be done with a dynamic hand off between sensors since the LEO orbital period is short compared to even the fast flyby time of a NEO within a lunar distance.

Space-based small satellites for planetary defense have been proposed recently \citep{2022Icar..37714906W}; though, not for all around characterization of discoveries made by the major surveys. I began by discussing the increasingly low cost of entry to LEO, and given the hole in our planetary defense characterization capabilities of our largest investments, it is clear that there is a need for a NEO characterization constellation in LEO. 

\section{Conclusions} \label{sec:conclusions}
In this paper, I broke down the NEO Survey Act into a series of ``missions'' that I used to assess the performance of the major NEO surveys at present (CSS and Pan-STARRS) as well as the two coming in the next few years (Rubin/LSST and NEO Surveyor). In Figure \ref{fig:9}, I detailed the gaps in the mission space left by the surveys, and in \S\ref{sec:followup}, I discussed the assets available to astronomers to help track and characterize NEOs after discovery. Finally in \S\ref{sec:constellation}, I discussed how a constellation of LEO small satellites would fill the holes left in particularly the NEO characterization mission. To conclude the paper, below is a list of the high level takeaways:
\begin{itemize}
    \item Despite a mandate to detect, track, catalogue, and characterize, NASA's investments have largely focused on NEO discovery and cataloguing. Tracking has been largely been implemented by astronomers (many who are amateur) that interface with the MPC.
    \item In 2004 when 99942 Apophis was discovered and determined to have a high chance of impact 25 years later, it took far too long to characterize the asteroid. Such a scenario is more likely when Rubin/LSST and NEO Surveyor are operational because they will detect far more NEOs than have been detected so far in a short period of time. 
    \item NEO threat landscape is more heavily weighted with objects not covered by the NEO Survey Act. There are $\sim$200 times more Chelyabinsk meteor and larger NEOs than there are 140 meter and larger NEOs.
    \item NEO characterization is best carried out by a combination of visible to NIR spectroscopy and resolved imagery with radar observations; however, these are far too expensive for the number of NEOs that have been discovered and impossible for the majority of NEOs due to their sub-optimal orbits and intrinsic faintness. Only about 15\% of NEOs have had a spectrum measured, and fewer than percent have been imaged with radar. This is in part because the majority of NEOs are too faint to observe with the ideal techniques for characterization. 
    \item In Figure \ref{fig:3}, I showed that the development of accurate spectral classifications with NEO spectroscopy can be used as a model to determine asteroid classifications with visible to NIR spectroscopy with obtainable photometric precision. 
    \item A worthy task for the planetary defense community would be to a set of visible to NIR filters that maximize the characterization potential of photometric measurements coving these wavelengths.
    \item Continued work with asteroid sample return missions and meteorite sample analysis have drawn credible links between the spectral classifications which further enables photometric classification to infer material properties. 
    \item Many NEO discoveries are impossible or difficult without decades long surveys due to the infrequency of their ventures into high Solar elongations. This is especially true for Atens and Atira asteroid (see Figure \ref{fig:6}). NEO Surveyor will be able to detect these asteroids; however, all current follow up instruments are ground-based, so NEO Surveyor detections will not be able to be characterized or tracked fully without a space-based option. 
    \item A space based constellation of small telescopes with visible and NIR sensors with several broad-band filters would enable the characterization of all discoveries made by NEO Surveyor and Rubin/LSST in near real time. 
\end{itemize}

\section*{Acknowledgements}
Thanks to Travis Yeager, Michael Schneider, Alex Geringer-Sameth, Megan Bruck Syal, Pete Supsinskis, Wim de Vries, Brian Bauman, and Ben Bahney for useful discussions leading to this paper. I also thank Kirsty Taggart, Connor Dickinson, and Srujan Dandu of UC Santa Cruz for generously observing the NEO that flew by Earth in December 2022 shown in Figures \ref{fig:1} and \ref{fig:10}. This work was performed under the auspices of the U.S. Department of Energy by Lawrence Livermore National Laboratory under Contract DE-AC52-07NA27344. Funding for this work was provided by LLNL Laboratory Directed Research and Development grant 23-ERD-044.

\bibliography{main}
\end{document}